\documentclass[useAMS,usenatbib]{mn2e}

\usepackage{graphicx}

\newcommand{\Msolyr}{~\mbox{M}_{\sun}\mbox{yr}^{-1}}
\newcommand{\Msol}{~\mbox{M}_{\sun}}
                     \newcommand{\yr}{~\mbox{yr}}
\newcommand{\Myr}{~\mbox{Myr}}
\newcommand{\Gyr}{~\mbox{Gyr}}
\newcommand{\GHz}{~\mbox{GHz}}
\newcommand{\keV}{~\mbox{keV}}
\newcommand{\Mpc}{~\mbox{Mpc}}

\newcommand{\cmet}{~\mbox{cm}}
\newcommand{\kpc}{~\mbox{kpc}}
\newcommand{\erg}{~\mbox{erg}}
\newcommand{\ergs}{~\mbox{erg~s}^{-1}}

\begin{document}

\title{Quenching Cluster Cooling Flows with Recurrent Hot Plasma Bubbles}

\author[C. Dalla Vecchia et al.]
{Claudio Dalla Vecchia,$^1$\thanks{claudio.dalla-vecchia@durham.ac.uk}
Richard G. Bower,$^1$ Tom Theuns,$^{1,2}$ Michael L. Balogh,$^1$ 
\and Pasquale Mazzotta,$^3$ and Carlos S. Frenk$^1$\\ \\
  $^1$Institute for Computational Cosmology, University of Durham, South Road, DH1 3LE Durham, UK\\
  $^2$University of Antwerp, Campus Drie Eiken, Pleinlaan 1, B-2610 Antwerpen, Belgium\\
  $^3$Dipartimento di Fisica, Universit\`a di Roma ``Tor Vergata'', via della Ricerca Scientifica 1, I-00133 Roma, Italy}

\maketitle

\abstract The observed cooling rate of hot gas in clusters is much
lower than that inferred from the gas density profiles.
This suggests that the gas is being heated by some source.
We use an adaptive-mesh refinement code ({\sc Flash}) to
simulate the effect of multiple, randomly positioned, injections of
thermal energy within 50~kpc of the centre of an initially isothermal
cluster with mass $M_{200}=3\times10^{14}\Msol$ and $kT=3.1$~keV.  We
have performed eight simulations with spherical bubbles of energy generated
every $10^8$ years, over a total of $1.5\Gyr$.  Each bubble is created
by injecting thermal energy steadily for $10^7$ years; the total
energy of each bubble ranges from 0.1--3$\times10^{60}\erg$, depending
on the simulation.  We find that 2$\times10^{60}\erg$ per bubble
(corresponding to a average power of $6.3\times10^{44}\ergs$)
effectively balances energy loss in the cluster and prevents the
accumulation of gas below $kT=1$~keV from exceeding the observational
limits.  This injection rate is comparable to the
radiated luminosity of the cluster, and the required energy and
periodic timescale of events are consistent with observations of
bubbles produced by central AGN in clusters.  The effectiveness of
this process depends primarily on the total amount of injected energy
and the initial location of the bubbles, but is relatively insensitive
to the exact duty cycle of events. \endabstract


\section{Introduction}

Gas cooling at the centre of a cluster halo is an inherently unstable process:
cooling increases the gas density, which in turn enhances the cooling
rate. X-ray observations show that the cooling time of gas in most
cluster cores is less than the Hubble time
\citep[e.g.][]{1977MNRAS.180..479F}. Unless cooling is balanced by
some form of heating, gas will flow into the cluster centre at rates
up to $\sim 1000\Msolyr$ \citep[e.g.][]{2001A&A...365L.104P}. What
happens to the cooled gas is unclear; it could be consumed by star
formation, or lead to a reservoir of low temperature ($<1\keV$)
material in the core. Although this may be the fate of a fraction of
the gas, there are indications that most gas in fact does not follow
either route.  First, the star formation rates in the central galaxies of
clusters are much lower than the inferred mass inflow rates
\citep{1989AJ.....98..180O,1987MNRAS.224...75J}, rarely approaching
$\sim 100 \Msolyr$. To estimate the star formation rate in a typical
cluster, we can average over the large sample of clusters studied by
\citet{1999MNRAS.306..857C}. This suggests that the typical star
formation rate is less than $10\Msolyr$.  Secondly, we can compare the mass
deposition rates with observations of the molecular gas content of
clusters.  \citet{2001MNRAS.328..762E} finds molecular gas masses
ranging from $10^9$ to $2\times 10^{11}\Msol$, with an average of
$2.6\times10^{10}\Msol$. Assuming a gas consumption timescale of
$10^9$ years \citep[see][]{2001MNRAS.328..762E}, this implies a
deposition rate that may be as high as $200\Msolyr$ in a few clusters,
but is $\sim 30\Msolyr$ on average.  Similar limits are obtained by
\citet{2003A&A...412..657S}. Finally, recent spectroscopic X-ray
observations show no evidence for significant gas cooling below
$1\keV$
\citep{2001A&A...365L..99K,2001A&A...365L.104P,2001A&A...365L..87T},
and observations of molecular and neutral material reveal
that the amount of cold gas in clusters of galaxies is also much less
than expected from the integrated cooling flow rate \citep[typically
less than $30\Msolyr$ -- ][]{2002MNRAS.337...49E,2003ApJ...594L..13E,2003A&A...412..657S}.

This cooling-flow paradox has led many authors to investigate
mechanisms to quench gas cooling. Observations of merging clusters
show little evidence for cooling flows, which suggests that the merger
process might be implicated in disrupting cooling.  Recent simulations
have shown that sub-halo merging can indeed heat up gas
\citep{2003astro.ph..9836B}; however, the amount of cold gas produced
in these simulations is still too large compared to that observed,
leading to the conclusion that additional heating processes must be
involved.  Several alternatives have been proposed, including energy
injection from radio sources or active galactic nuclei
\citep[AGN;][]{1995MNRAS.276..663B,2001MNRAS.328.1091Q,2002MNRAS.332..729C},
viscous dissipation of sound waves
\citep{2003MNRAS.344L..43F,2003astro.ph.10760R} and thermal conduction
\citep{2003astro.ph..8352V,2004astro.ph..1470D}.  Each of these has
advantages and disadvantages.  In particular, it is difficult to
balance cooling, with its $\rho^2$ density dependence, with heating
processes that typically scale as $\rho$.  Since cooling and
heating can then balance only at one particular density, some of these
feedback mechanisms may lead to unstable solutions with some regions
of the cluster being efficiently heated while others continue to cool
catastrophically.

AGN are particularly promising candidates for balancing cooling, given
their potentially large energy reservoir
\citep{1993MNRAS.263..323T,1995MNRAS.276..663B,2001MNRAS.325..497B}.
In most numerical simulations of this process to date, energy
injection produces bubbles at high temperature and low density that,
after a short expansion phase, gain momentum by buoyancy
\citep{2001MNRAS.325..676B,2001MNRAS.328.1091Q,2003MNRAS.339..353B}.
This mimics the effect of a jet which rapidly loses its collimation,
as often observed in local clusters \citep{2003astro.ph.10011E}.  In
other simulations, gas is injected at high velocity to mimic jets
which retain their large-scale coherence
\citep{2002MNRAS.332..271R,2003astro.ph..7471O}.  In this paper, we
consider the first mechanism and show that such bursts of localised energy
can induce convection of the intra-cluster medium (ICM), which leads
to a quasi-stable cluster configuration and reduces the average mass
deposition rate to within observational limits. We
discuss whether the required energy and duty cycle of AGN
activity are compatible with observational limits.  This paper is
organised as follows. In Section~\ref{sec:simul} we describe the setup
of our simulations and discuss the parameters explored. Results are
presented in Section~\ref{sec:results} and, in
Section~\ref{sec:discuss}, we compare the required energy and duty
cycle with observational constraints. Finally, our conclusions are
summarised in Section~\ref{sec:conc}.

\begin{table}
  \begin{center}
    \begin{tabular}{ccccc}
      \hline
      & & \multicolumn{2}{c}{\sc Injected/Radiated Power} \\
      name  & $E$           & $\dot E_i$      & $<\dot E_r>_{100}$ \\
            & $10^{60}\erg$ & $10^{44}\ergs$  & $10^{44}\ergs$ \\
      \hline
      A     & 0             & 0               & no cooling \\
      \hline
      S0.0  & 0             & 0               & 28 \\
      \hline
      S0.1  & 0.1           & 0.32            & 18 \\
      \hline
      S0.3  & 0.3           & 0.95            & 18 \\
      \hline
      S0.6  & 0.6           & 1.9             & 13 \\
      \hline
      S1.0  & 1.0           & 3.2             & 13 \\
      \hline
      S1.5  & 1.5           & 4.7             & 8.7 \\
      \hline
      S2.0  & 2.0           & 6.3             & 7.9 \\
      \hline
      S3.0  & 3.0           & 9.5             & 6.8 \\
      \hline
    \end{tabular}
  \end{center}
  \caption{Summary of the nine simulations performed. The different
    simulations are referred to as S$n$, where the energy $E$ of a
    single bubble is $n\times 10^{60}$ erg. $\dot E_i$ is the mean
    energy injection rate over a duty cycle, and $<\dot E_r>_{100}$ is
    the mean emitted energy rate, averaged over the last $10^8\yr$ of
    the simulation.}
  \label{tbl:inject}
\end{table}

\begin{figure*}
\includegraphics[width=41pc]{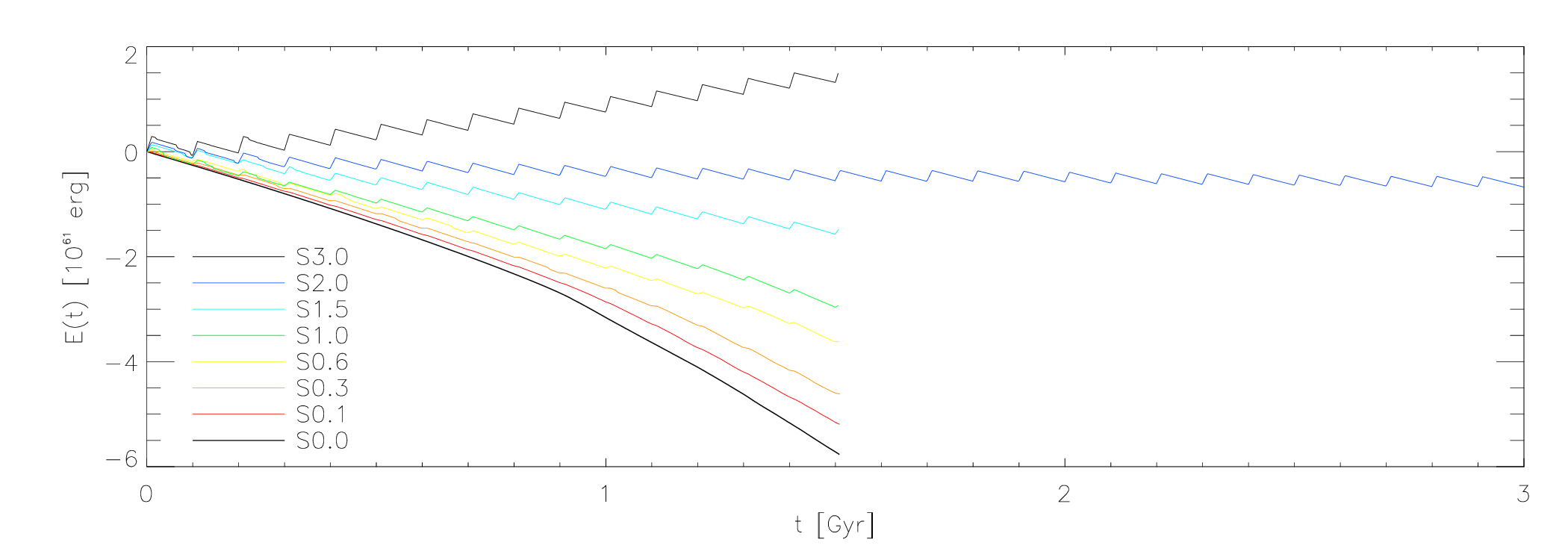}
\caption{The evolution of the total energy of each simulation is shown
  by plotting $\Delta E(t)=E_{\rm T}(t) - E_{\rm T}(0)$, where $E_{\rm
  T}(t)$ is the sum of internal, kinetic and potential energy at time
  $t$ and $E_T(0)$ is its initial value.  The saw-tooth shape of the
  curves results from the discrete AGN events and subsequent cooling.
  At the mean injection rate of simulation S2.0, the energy keeps an
  almost constant value within the simulation time. We tested that
  this behaviour is maintained up to $5\Gyr$, though we show the
  evolution only to $3\Gyr$ for clarity.}
\label{fig:energy}
\end{figure*}

\section{Simulations}

\label{sec:simul}

In order to concentrate our numerical resolution on the interactions
between the cooling material and the buoyant hot bubbles, our simulations
use a fixed external gravitational potential and neglect the self-gravity of 
the gas. This allows us to run the simulation over cosmologically
significant timescales. We model the input of energy as a cycle of 
energy injection and quiescent phases. In the active phase a small 
region of the gas is heated so that it expands to form a bubble that 
rises out of the confining potential. 

\subsection{The code}

Our simulations were performed with {\sc Flash}~2.2, an Adaptive Mesh
Refinement (AMR) hydrodynamical code developed and made public by the
ASCII Center at the University of Chicago
\citep{2000ApJS..131..273F}. {\sc Flash} is a modular block structured
AMR code, parallelised using the Message Passing Interface (MPI)
library.

{\sc Flash} solves the Riemann problem on a Cartesian grid using the
Piecewise-Parabolic Method \citep[PPM;][]{WC}. It uses a criterion based on the
dimensionless second derivative ${\cal D}^2\equiv |(F
d^2F/dx^2)/(dF/dx)^2|$ of a fluid variable $F$ to increase
the resolution adaptively whenever ${\cal D}^2>c_2$ and de-refine the grid when
${\cal D}^2<c_1$, where $c_{1,2}$ are tolerance parameters. When a
region requires refining (${\cal D}^2>c_2$), child grids with cell size
half that of the parent grid are placed over the offending region, and
the coarse solution is interpolated. In our simulations, we have used
density and temperature as the refinement fluid variable $F$
\citep[see][for more details]{2000ApJS..131..273F}.  We
have performed a series of simulations with a different maximum level
of refinement to investigate the effect of numerical resolution, as
described in Section~\ref{sec:sims}

{\sc Flash} interpolates the imposed analytic gravitational potential
(see Section~\ref{sec:ic}) to the grid cells, and computes the
corresponding gravitational acceleration, neglecting the self-gravity of the gas.  To interpolate the
initial density field from our analytical solution to the {\sc Flash}
mesh, we impose an initial grid with increasing resolution from large
radii to the centre of the cluster. This is the minimum refinement
level allowed during the simulation.

For the boundary conditions, we impose that the values of density,
temperature and velocity on the boundary cells (guard cells) remain
the ones computed at the initial time and satisfy the hydrostatic
equilibrium conditions.  With the cluster located at the centre of the box, the
box size of $6\times 10^{24}\cmet$ ($1.9\Mpc$) is large enough to
ensure that cooling does not affect the boundaries.  This guarantees
the hydrostatic equilibrium condition is not broken, and no evident
inflow or outflow occurs at the border of the grid. Moreover, the
temperature difference between the last computational cell and the
guard cell next to it, at the end of the simulation, is always less
that 2\%.  Likewise, sound waves propagating across the volume do not
generate any unphysical artifacts at the boundary.

The time step is derived by the Courant condition ${\rm d}t = C \Delta x/c_s$, where $\Delta x$ is the dimension of a cell, $c_s$ is the
sound speed in that cell and $C$ is a coefficient, usually
less than one. The time step is uniform; thus, all the cells evolve at
each time step.

\subsection{Initial Conditions}\label{sec:ic}

The time-independent gravitational potential of the dark matter is that
corresponding to a \citet{1996ApJ...462..563N,1997ApJ...490..493N} 
density profile

\[
  \rho_{\rm DM}/\rho_{\rm crit}=\frac{
    \delta_{\rm c}}{(r/r_{\rm s})(1+r/r_{\rm s})^2}\,,
\]
where $\rho_{\rm crit}$ is the critical density, $r_{\rm s}=r_{200}/c$
and $r_{200}=1.38\Mpc$. We adopted a halo mass of 
$M_{200}=3\times 10^{14}\Msol$ and set the concentration parameter to
$c=6$ \citep{2001ApJ...554..114E,2002ApJ...568...52W,2004astro.ph..1470D}.

We use the prescription of \citet{2000MNRAS.318..889W} to set up an
isothermal gas distribution in hydrostatic equilibrium within this
dark matter potential.  We adopted this prescription because of its
simplicity. Fits to numerical simulations \citep{2001MNRAS.327.1353K},
suggest that the temperature should decline with increasing radius.
However, our isothermal profiles do match the observed temperature
structure of clusters within $0.2 r_{200}$, where the observational
data show an isothermal temperature profile
\citep[eg.][]{2001MNRAS.328L..37A,2001ApJ...551..153D,pointecouteau}
sometimes with a central decrement.  A central decrement is quickly
generated as the model cluster cools.

We choose an initial gas temperature of $T=3.1\keV$, consistent with
the observed correlation between temperature and $M_{200}$
\citep{2002ApJ...567..716R}.  The appropriate gas density is then
\[
  \rho_{\rm GAS}/\rho_0=(1+r/r_{\rm s})^{\eta/(r/r_{\rm s})},
\]
where $\rho_0$ was chosen to satisfy the relation
\[
  M_{\rm GAS}^{200}/M_{\rm DM}^{200}=\Omega_b/(\Omega - \Omega_b),
\]
Here, $M^{200}_{\rm DM}$ denotes dark matter mass contained within a
sphere whose mean interior mass density is 200 times the critical
density.  We used $\eta=10.25$, $\Omega=0.3$ and $\Omega_b=0.04$ for
the matter and baryon densities in units of $\rho_{\rm crit}$, and
$h=0.7$ for the Hubble parameter.

We adopt the cooling functions of \citet{2002ApJ...567L.103T}, which
include cooling by H, He and metals, in the presence of an ionising
UV-background. This cooling function uses interpolation tables for the
cooling rate due to a solar admixture of metals, obtained from {\sc
Cloudy} \citep[version 94,][]{1998PASP..110..761F}.  We assume a
metallicity of one third solar, and the \citet{1996ApJ...461...20H}
UV-background at redshift $z=0$, as updated by
\citet{2001cghr.confE..64H}. The cooling routine was tested against
the {\sc MEKAL} \citep{1986A&AS...65..511M} cooling routines in the
X-ray analysis package {\sc XSPEC} \citep{K92} and against the cooling
rates used by \citet{MCI}.

Heating by an AGN is modelled by injecting energy at predefined points
in space and time. This allows us to simulate multiple episodes of AGN
activity as we now describe.

\subsection{Energy Injection}

To simulate the effect of a central AGN, bubbles of energy are
distributed randomly inside a radius of $50\kpc$ from the centre.  We
do not attempt to model the processes leading to the formation
of the bubbles, but we consider it likely that the bubbles will be 
injected at different positions because of the precessional motion of the
jet axis of the central source.

Energy is injected with a three-dimensional Gaussian distribution
proportional to $e^{(x^2+y^2+z^2)/2\sigma^2}$ and we take the
initial size of the bubble to be characterised by $\sigma=10.3$ kpc.
The bubble is truncated at a maximum radius of
$r_{\rm b}=5\sigma=51.5\kpc$, which is seven cells
in our standard six level resolution runs.  The final radius
(after the expansion of the bubble) depends on the amount of energy
injected.  

Starting from $t=0$, energy is injected every $10^8\yr$ for
a time of $10^7\yr$ in our standard simulations. We assume energy
values from $1\times 10^{59}$ to $3\times 10^{60}\erg$
(Table~\ref{tbl:inject}).  Our choice of parameters deliberately lies
at the upper end of observed AGN duty cycles and jet powers
\citep{1999MNRAS.309.1017W,2000ApJ...543..611O,2003MNRAS.344L..43F,
2003astro.ph.10011E}.

\begin{figure*}
\includegraphics[width=41pc]{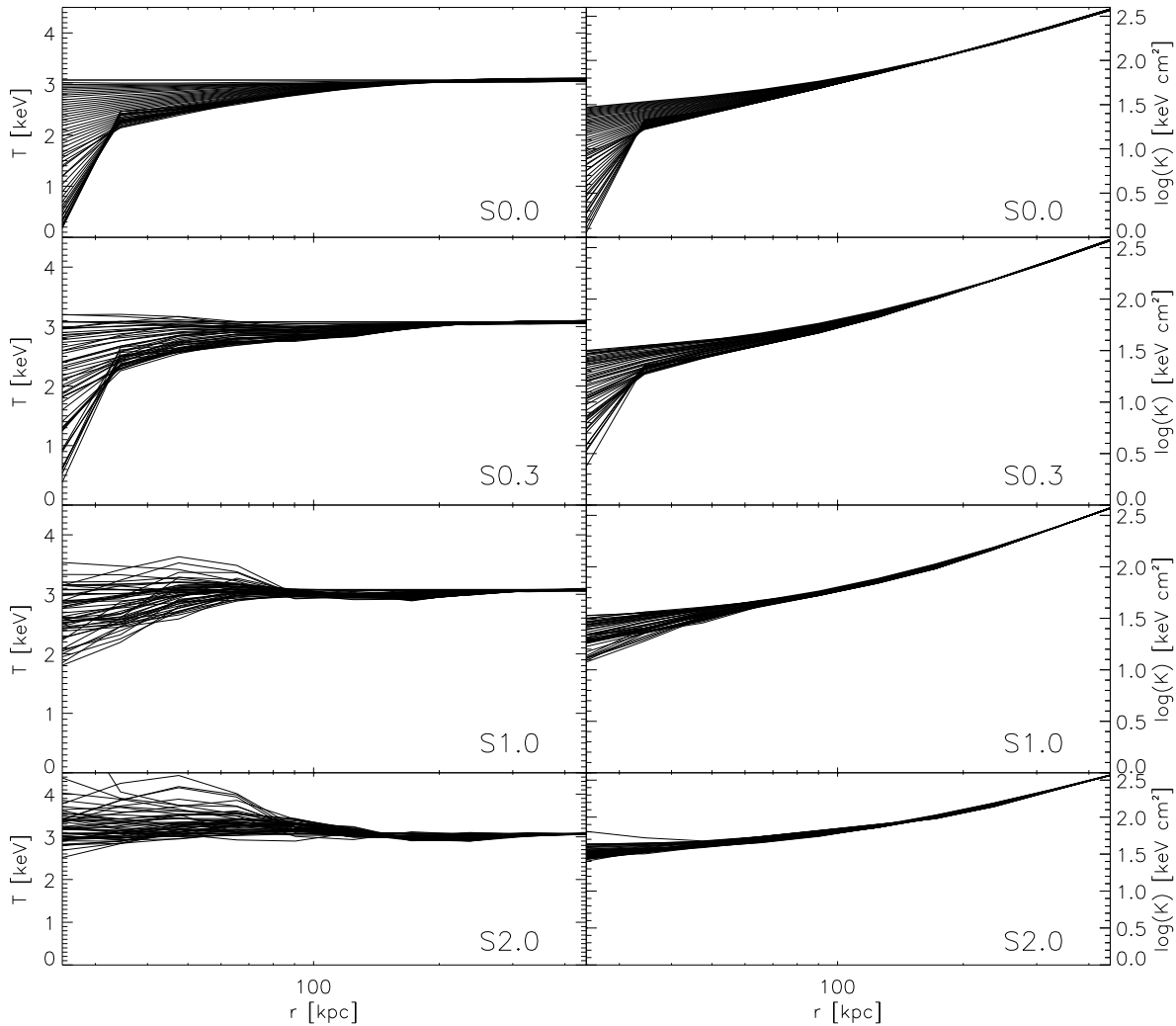}
\caption{The emissivity--weighted temperature (left panels) and
  entropy (right panels) profiles of each simulated cluster are shown
  for 50 simulation times. From top to bottom,
  simulations S0.0 (without any AGN bubble), S0.3, S1.0 and S2.0 are
  shown. The effect of energy injection is to maintain the
  initial, relatively shallow entropy profile and isothermal
  temperature in the cluster core.}
\label{fig:temp_entr}
\end{figure*}

\subsection{The Simulations}\label{sec:sims}

Our main results are derived from a series of nine simulations
(Table~\ref{tbl:inject}):
\begin{itemize}
\item[1.] an adiabatic test simulation (A) without cooling or AGN bubbles;
\item[2.] a simulation with cooling but no AGN bubbles (S0.0);
\item[3.] seven simulations with cooling and AGN bubbles of varying
  strength (S0.1--S3.0), but the same duty cycle. In these simulations
  only the energy of each bubble varies. The position and timing of
  the heating events are the same.
\end{itemize}
The results of these simulations are presented in Section~\ref{sec:results}.

We also performed simulations with the following changes in order to
test the sensitivity of our results to the details of the heating cycle:
\begin{itemize}
\item[4.] the bubbles were randomly distributed inside a sphere of 25
  or 100~kpc, instead of 50~kpc;
\item[5.] the same total number of bubbles (15) was generated within
  1.5~Gyr, but at random time intervals; 
\item[6.] the energy was injected as bubble pairs, with each bubble
  containing half the energy per event.
\end{itemize}
Finally, we also reran simulation S2.0 with increased resolution to
check for numerical convergence.  The results of this test, described
in Appendix~\ref{sec:conv}, led us to adopt a maximum refinement level
of six. We also continued the simulation S2.0 to 5~Gyr to test the
long term stability of this simulated cluster.

In the adiabatic simulation, the gas remains almost static as expected
for our equilibrium set-up, apart from small re-adjustments induced by
discretisation. Energy and mass are conserved to $\simeq 0.4\%$ and
1\% respectively.
With cooling allowed (simulation S0.0), the loss of pressure support
causes a cooling flow to be established. 

The intial conditions have a relatively high X-ray luminosity ($\log
L_{X,bol} = 45.0$) compared to the observed X-ray luminosities of
clusters of this mass and temperature \citep{1998ApJ...504...27M,MCI}.
This is a well-known problem that results from the low entropy gas in
the centre of a cuspy dark matter potential.  Starting with such a
luminous cluster ensures that our simulations conservatively explore
the maximum energy necessary to stabilise cooling.  In order to be
fully successful, however, our energy injection must not only balance
the cooling, but also raise the entropy of the cluster gas
sufficiently to reduce the total luminosity by a factor $\sim 10$.

By default, we ran each simulation for $1.5\Gyr$. Without a cosmological
context, a longer simulation is not well justified
since cluster-cluster mergers provide an important additional mechanism 
for re-organising the gas distribution and erasing the cooling
flow structure.  Nevertheless, we ran the S2.0 simulation for $5\Gyr$ to 
check the evolution of the solution over very long time scales.

\section{Results}

\label{sec:results}

\subsection{The Effect of Energy Injection}

In Figure~\ref{fig:energy} we show, for each simulation, the variation in the
total energy within the simulation volume. The pulsed
injection of the energy and subsequent cooling gives rise to a saw-tooth
pattern.

In simulation S0.0, which has no heating, the cooling rate increases
with time as the cluster collapses. Including energy injection
(simulations S0.1--S3.0) reduces the overall amount of energy that is
radiated.  This is not a trivial result since the energy injection
events can actually promote cooling in the compressed regions around
the rising bubbles.  In practice, however, the dominant effect is the
mixing of high and low entropy material in the large-scale gas motions
induced by the bubbles.  Therefore, as the injected energy is
increased, the amount of radiated energy drops.  At a mean energy
injection rate of $6.3\times 10^{44}\ergs$ (simulation S2.0), the
radiated energy and the injected energy are balanced.  This
balance does not guarantee that the structure of the cluster is
stable, as continuing collapse of the central region might be balanced by
expansion of the outer part of the cluster.  However, as
Figure~\ref{fig:temp_entr} shows, the whole entropy profile of the
cluster changes little.  We investigated the long term behaviour of
this model by continuing this simulation for $5\Gyr$
(Figure~\ref{fig:energy}) and found that the cluster maintains this
behaviour even over this longer timescale. Of course, over such long 
timescales, the effects of cluster mergers can not be neglected.

\begin{figure}
\includegraphics[width=20pc]{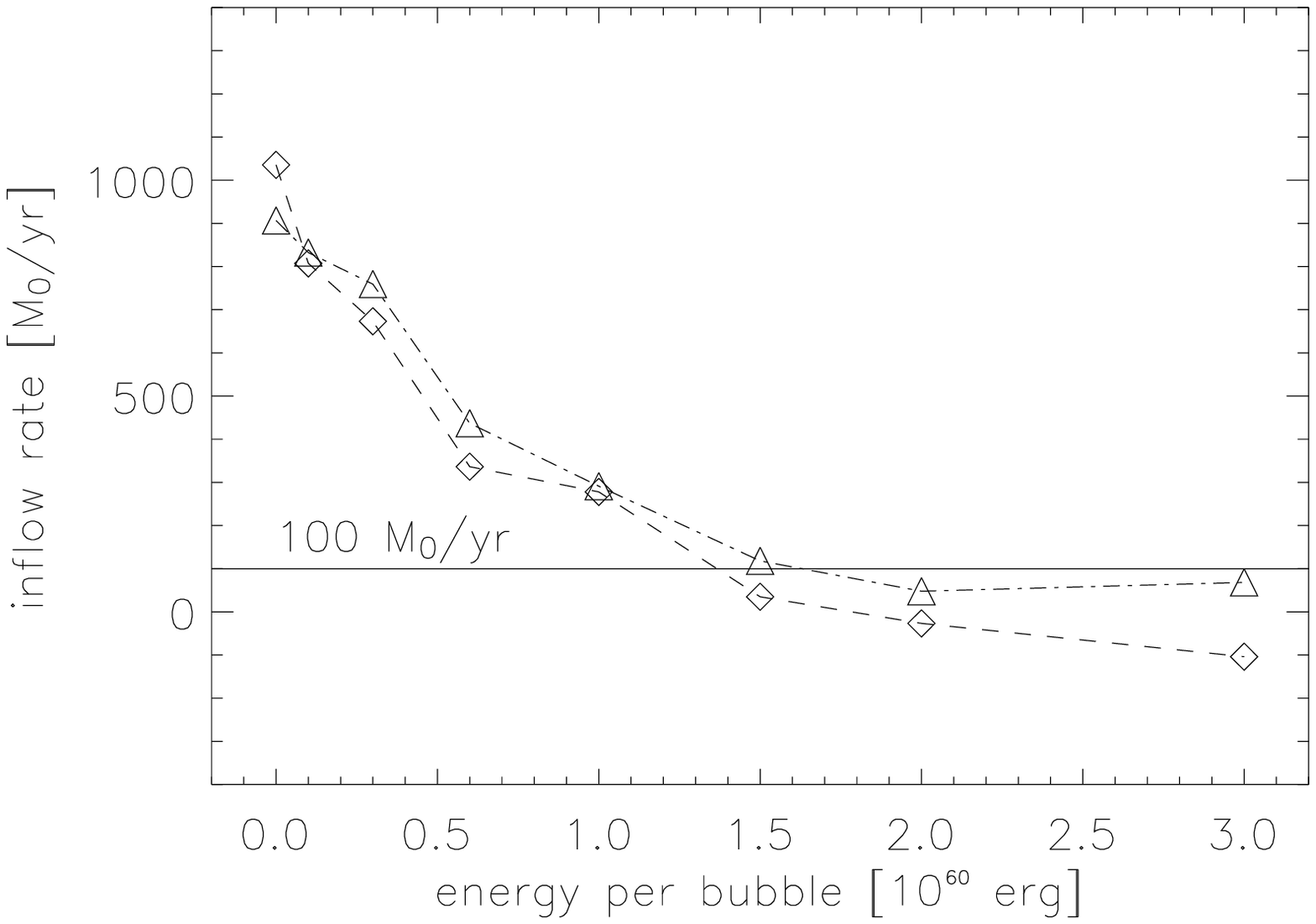}\\
\includegraphics[width=20pc]{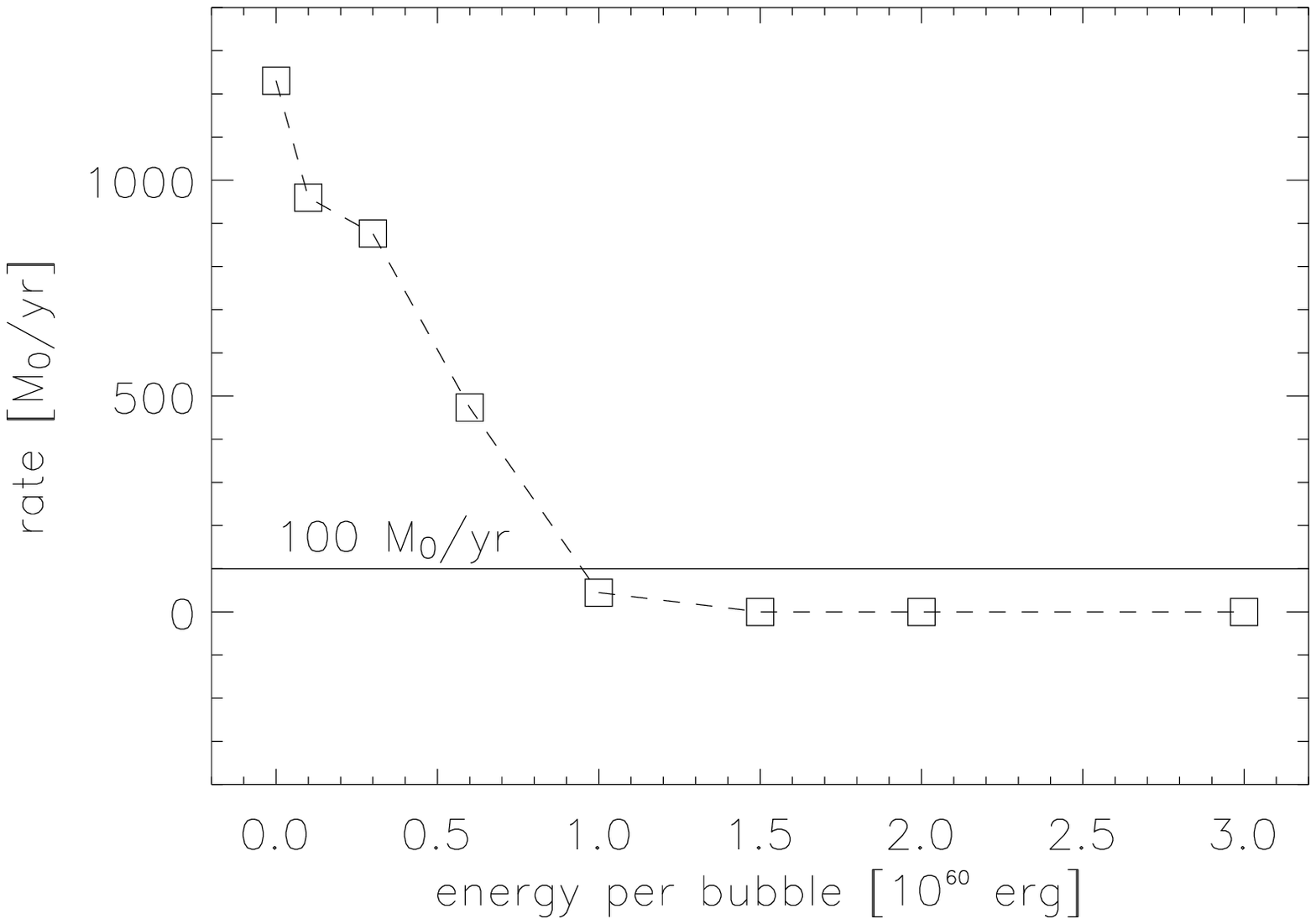}
\caption{\textbf{Top panel}: The mass deposition rate into a sphere of
  radius $50\kpc$ (dashed line) and $100\kpc$ (dash dotted line),
  averaged over the time interval $1.2$ to $1.5\Gyr$, as a function of
  the energy injected per bubble. For bubbles of energy above
  $10^{60}\erg$ the deposition rate is $\la 100\Msolyr$.
  \textbf{Bottom panel}: The mean growth rate of the total mass at
  temperature $<1\keV$ for each simulation. The amount of gas cooling
  below $1\keV$ drops down to almost zero when the injected energy per
  bubble is greater than $10^{60}\erg$.}
\label{fig:rates}
\end{figure}

The emissivity-weighted temperature and entropy profiles of each
simulation are shown, for 50 output times, in
Figure~\ref{fig:temp_entr}.  We define the entropy as $K\equiv
kT_en_e^{-2/3}$ [keV~cm$^2$], where $T_e$ and $n_e$ are the electron
temperature and density, respectively.  This is related to the
thermodynamic entropy by a logarithm and additive constant.  If
cooling is not balanced by the energy injection, the entropy profile
drops dramatically in the centre (top-right plot).  This is reflected
as a marked dip in the temperature profile of the cluster (top-left
plot).  Energy injection reduces this trend, maintaining the
relatively flat entropy profile and isothermal temperature of the
initial cluster.

The cooling of material in the central cluster regions leads to a net
inflow of gas.  In Figure~\ref{fig:rates}, we plot the mass
accumulation rate within the central regions of our simulated
clusters.  The rate is determined by averaging over the last $0.3\Gyr$
of the simulation, and is computed within the central $100\kpc$ and
$50\kpc$.  The results for both radii are similar, as expected if the
cluster is in a quasi-steady state.  The figure also shows the mass
accumulation rate of material with temperatures less than $1\keV$.  In the
absence of energy injection, the cooling cluster deposits material in
the cluster centre at a rate that exceeds $1000\Msolyr$, a factor
$\sim 10$--$100$ higher than the observational limits
\citep{1999MNRAS.306..857C,2001MNRAS.328..762E,2003A&A...412..657S}.
Similarly, this model has a large mass flux of material cooling below
$1\keV$ ($1/3$ of the ambient temperature).

Energy input at an average rate of $4.7\times10^{44}\ergs$ (S1.5)
reduces the deposition rate below $30\Msolyr$ at the end of the
simulation, as required by observations
\citep{2001A&A...365L.104P,2001A&A...365L..87T,2001A&A...365L..99K,2003ApJ...590..207P}.
Even with a lower energy injection rate of $3.2\times10^{44}\ergs$
(S1.0), the rate at which material cools below $1 \keV$ is still
consistent with observed X-ray spectra.  We conclude that energy
injection through the creation of buoyant bubbles can successfully
prevent catastrophic cooling.

\begin{figure*}
\begin{center}
\includegraphics[width=17.5cm]{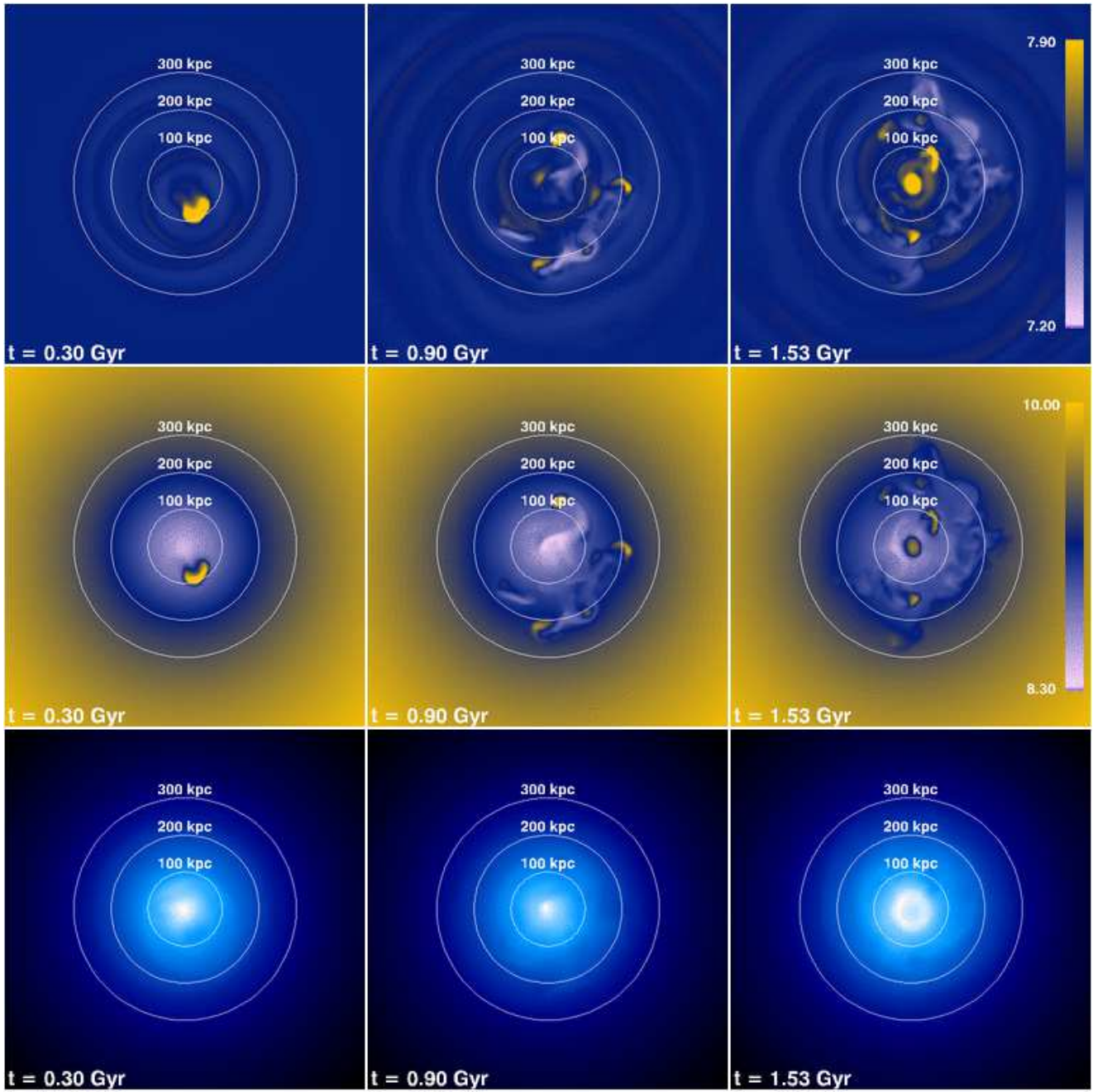}
\end{center}
\caption{The time evolution (from left to right) of simulation
  S2.0. From top to bottom, the quantities shown are the temperature
  [K] and entropy [ergs g$^{-2/3}$ cm$^2$] on the x-y plane crossing
  the centre. In the bottom row we show the approximate bolometric
  emissivity $\rho^2T^{1/2}$ projected through the simulation
  volume. The temperature distribution reveals the presence of sound
  waves propagating through the ICM. The sound waves are almost
  concentric and regular, a consequence of the periodic energy
  injection events near the cluster centre.}
\label{fig:S2.0}
\end{figure*}

\subsection{Two-dimensional morphologies}

In Figure~\ref{fig:S2.0}, we show the time evolution of various
quantities in two-dimensional sections of simulation S2.0.  The
temperature and entropy distributions are shown for a slice in the
x-y plane. They show that cold, low entropy gas is
pushed to large radii by the bubble's buoyancy and mixed with gas at
those radii. Thus, the cooling rate is reduced not by direct heating
of the cooling gas, but by convective transport of this material to
regions of lower pressure.  Similar results have been seen in
simulations of single bubbles
\citep[e.g.][]{2001MNRAS.328.1091Q,2002MNRAS.332..729C}.

The temperature distribution also reveals the presence of sound waves
propagating through the ICM. The sound waves are almost concentric and
regular, a consequence of the periodic energy injection events near
the cluster centre.

\begin{figure}
\includegraphics[width=20.5pc]{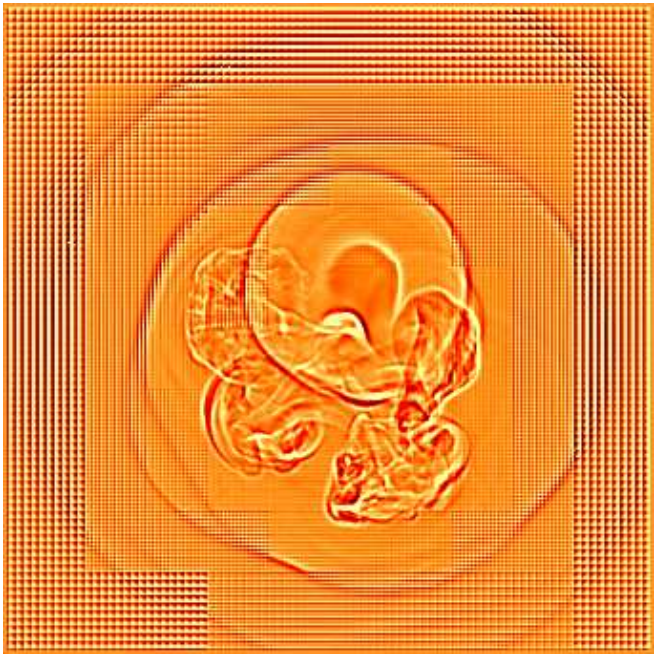}
\caption{An unsharp masked projection of the approximated X-ray
  emissivity, $\rho^2 T^{1/2}$, for the central ($0.9\Mpc$ diameter)
  part of the cluster simulated at a maximum resolution of $\sim
  2\kpc$.  The ``granular'' structure of the image is due to the
  resolution of the simulation grid.  The ripples from successive
  energy injection events can be clearly seen.}
\label{fig:xray}
\end{figure}

In order to illustrate what might be observable with an idealised
 X-ray satellite,
we also calculated the projection of the quantity $\rho^2T^{1/2}$, which is
approximately proportional to the bolometric gas emissivity.  Despite
the large amounts of energy being dumped into the central regions, the
large-scale gas distribution appears smooth and undisturbed in
projection.  To reveal the presence of perturbations on this smooth
profile, we created an unsharp masked image through the transformation
$(\mbox{img}_0-\mbox{img}_{\rm s})/\mbox{img}_{\rm s}$, where
$\mbox{img}_0$ and $\mbox{img}_{\rm s}$ are the original image and its
smoothed version, respectively.  The initial image was taken from a
high resolution simulation ($\sim 2\kpc$) and the
final unsharped image is shown in Figure~\ref{fig:xray}. The contrast
in this image is high enough to show the sound waves distinctly as
\textit{ripples}.  
As suggested by \citet{2003MNRAS.344L..43F}, these ripples in X-ray
images of observed clusters could be the fingerprint of multiple
outbursts and could be used to measure their periodicity.
Furthermore, if the cluster gas is sufficiently viscous, these sound
waves might help to offset the cooling flow by directly heating the
gas \citep{2003MNRAS.344L..43F,2003astro.ph.10760R}. By comparing the
properties of successive ripples it may be possible to measure the
viscosity of the ICM.  
Although the ripples in our simulation extend to very large radii, a result 
of the small viscosity in the simulations, it is unlikely that realistic
observations would have sufficient signal--to--noise ratios to see these
weak features at such distances.
We will investigate this further in a future
paper, where we synthesise artificial X-ray images, with the
appropriate transmission functions, from our simulations.

\subsection{Sensitivity to duty cycle parameters}\label{sec:var}

Simulations S0.1--S3.0 were based on a regular injection of
energy. This is somewhat artificial, so we investigated whether
randomly--timed energy injection events would have the same effect.
We also investigated what happens if the energy injection occurs in
less frequent bursts of greater intensity, or if bubbles are
distributed inside a sphere of larger ($100\kpc$ instead of $50\kpc$)
or smaller ($25\kpc$) radius. All our tests were performed choosing the
parameters of simulation S2.0.

\begin{figure*}
\includegraphics[width=41pc]{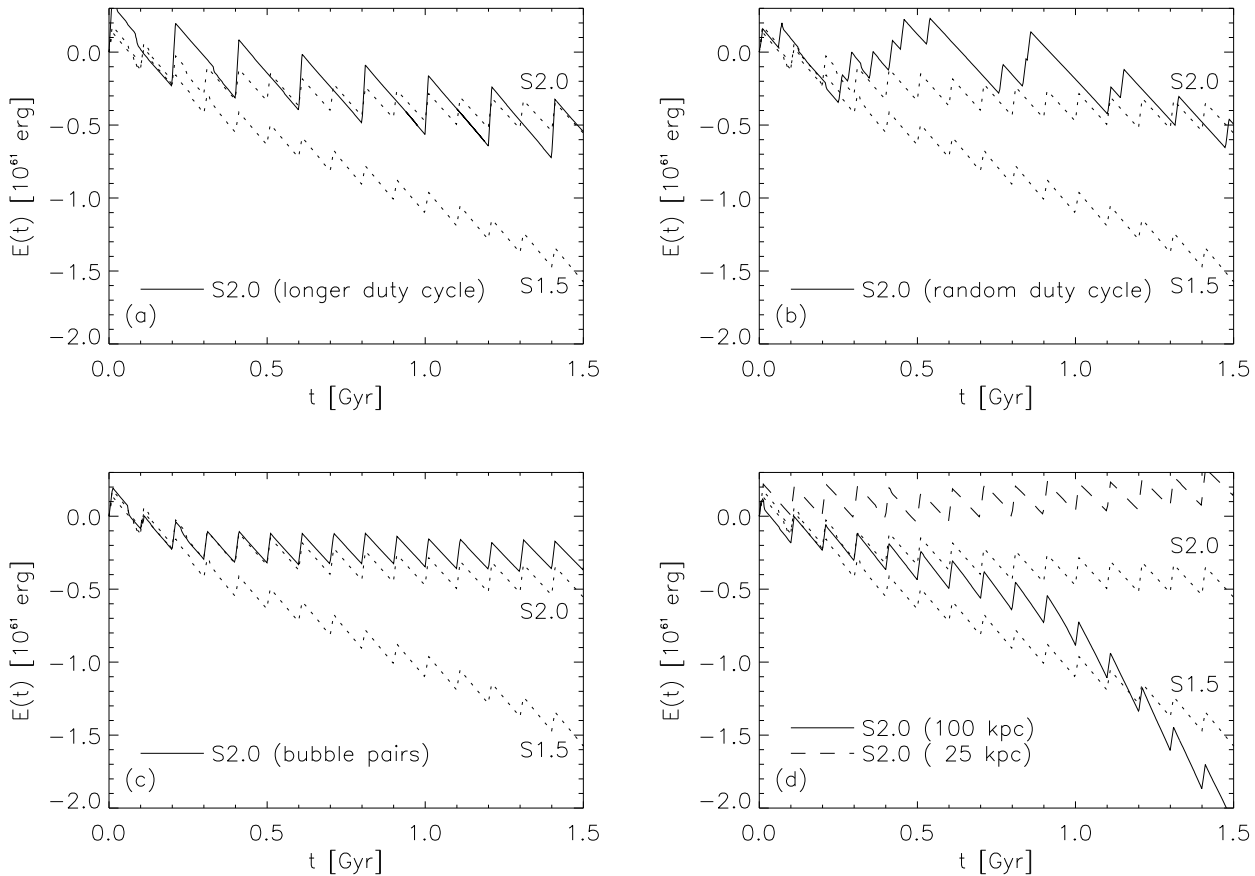}
\caption{In each panel we show the evolution of the total energy in
  simulations S2.0 and S1.5 (dotted lines), as in
  Figure~\ref{fig:energy}.  The other lines represent models with the
  same total energy as simulation S2.0, but with different parameters
  for the energy injection mechanism.
  In panel (a) we show a simulation where the duty cycle is
  regular, but twice as long as in S2.0.
  In panel (b) we show the
  effect of randomising the time interval between injection events.
  In panel (c) we
  show the effect of injecting the energy within pairs of bubbles. 
  Finally, in panel (d) we show, as the
  dashed and continuous
  curves, the effect of placing the bubbles randomly within
  a 25 kpc or 100 kpc area, respectively.  }
\label{fig:params}
\end{figure*}

The resulting energy evolution of each simulation is illustrated in
Figure~\ref{fig:params}.
First, we double the time between bursts,
while keeping the total average energy rate constant; thus, each
bubble has twice as much energy injection as in S2.0.  In this case,
the variation in total energy during each cycle is much larger than
before.  However, the long term evolution is very similar. Randomising
the injection time does not have a large effect on the average energy
evolution over the duration of the simulation either.  Where a
particularly long time interval elapses between bubbles, the energy
drops further as cooling progresses; however this is offset by later
events where bubbles appear in quick succession.  Our results are also
insensitive to whether the bubbles are injected singly, or in pairs
with the total energy divided between the two bubbles, possibly a
better model for the AGN activity.  The main factor that does make a
significant difference is the location of the bubbles.  Placing the
bubbles initially within a larger radius (100 kpc) is much less
effective at disrupting the cooling flow, since these bubbles tend to
rise buoyantly without disturbing the central, cooling gas.  By the
end of this simulation, the total cluster energy has dropped below
that of simulation S1.5 and is falling rapidly.  Conversely, placing
all the bubbles within 25 kpc is much more effective at preventing
cooling, and the total energy of the system actually increases with
time. To summarise, we see that the energy balance is most sensitive
to the total amount of injected energy, and to the initial location of
the bubbles.

\section{Discussion}

\label{sec:discuss}

\subsection{Energy Requirements}

Most central galaxies in clusters are radio sources
\citep{1996AJ....112....9L}; we can therefore expect that events like
those we have simulated will play an important role in cluster
evolution.  We have shown that the cluster cooling rate can be reduced
so that it agrees with observations if the AGN activity is
sufficiently energetic.  We have modelled the energy injection events
as short outbursts separated by relatively long quiescent periods in
order to make an extreme test of the long term effect of the energy
injection mechanism. To stabilise the long term evolution of the model
cluster in this way, we require an energy of $\sim 2\times 10^{60}
\erg$ for each of the injection events, if the bubbles are distributed
within a $\sim 50$ kpc radius.  We can associate these events with
outbursts of powerful AGN activity with a duration of $\sim 10^7 \yr$
and a power of $\sim 6\times 10^{45} \ergs$.

The observed radio power of central galaxies varies by factors of
$\sim100$ for clusters with similar core X-ray luminosities.  The most
powerful radio sources can have monochromatic radio power of
$\nu\cdot S_{\nu} \sim 10^{41.5}\ergs$ (where $S_{\nu}$ is the radio
luminosity density at $\nu =1.4$\GHz); around $10\%$ of clusters host
such powerful sources.  Eilek's (2003) analysis of central cluster
radio source morphologies and ages suggests that the ratio of the
total jet power ($P_{\rm jet}$) to $\nu\cdot S_{\nu}$ is more than
$10^4$. Under minimal assumptions, this can be confirmed in the case
of M87, where the most detailed radio observations are available
\citep{2000ApJ...543..611O}; such powerful jets are also inferred for
classical FR II sources \citep{1999MNRAS.309.1017W}.  This power is
close to the Eddington limit for accretion by a $10^9 \Msol$ black hole.
Thus, the total jet power of the most powerful radio galaxies is
comparable to the energy required to counter-balance cooling in our 
simulated cluster.
The frequency with which these sources occur in clusters also seems
consistent with the duty cycle that we have assumed in our
simulations.

It is worthwhile to emphasise that the energy injection can vary 
significantly (by a $\sim 50\%$) while still reducing the mass
deposition rate to an acceptable level after 1.5 Gyr. These variations
in the injected energy lead to an overall cooling or heating of 
the cluster: however subsequent mergers with massive substructures
will disrupt the evolution presented in our simulations (which assume
a static gravitational potential). The shocks and turbulence generated 
in the mergers will lead to a re-distribution of energy and entropy
that may erase the differences that have built up during the quiescent 
phase that we simulate
\citep{2002ApJ...569..122G,2003astro.ph..9836B}.
Our model does not contain any mechanism to regulate the
energy injection -- this would require detailed understanding of the 
physics powering the central source from the surrounding gas reservoir.
However, as 
\citet{1995MNRAS.276..663B} \citep[see also][]{2004MNRAS.347.1093B}
have argued, it seems natural that the cooling rate and the energy injected 
by the central energy source should be linked. In this case, a balance
between heating and cooling is naturally expected when these quantities are
integrated over sufficiently long timescales: if insufficient energy is
injected into the ICM, the net cooling leads to a high mass deposition 
rate, and thus an increase in the fueling of the central engine. Nevertheless,
the delays
inherent in this feedback loop lead to a short-term imbalance, and thus to
the out bursts that we model here.

\subsection{Global X-ray properties}
Although the injected energy is able to counterbalance the cooling flow in
the cluster, the effect on the cluster's overall structure is modest,
as we showed in Figure~\ref{fig:temp_entr}.  In the absence of any
heating, cooling causes the central entropy and temperature to drop
precipitously.  In simulation S1.0, the AGN energy is sufficient to
reduce this drop to only $\sim 50$ per cent.  Thus, by preventing
runaway cooling, our energy injection mechanism reduces the change in
the initial entropy profile of the cluster.  The result is that the
integrated luminosity and average temperature of the cluster evolve little.
Doubling the amount of energy injected (S2.0) results in a small
increase in the central temperature ($\sim30$\%), and a 
20\% reduction in luminosity.

We simulated an initially isothermal cluster,
with temperature observationally consistent
with its mass \citep{2002ApJ...567..716R}, in order to see how it would
evlove in the luminosity-temperature plane. The experiment raises two
issues that are not fully resolved by our model: (1)~The cluster is too
luminous, for its emission weighted temperature, compared with observed
clusters \citep[e.g.,][]{1998ApJ...504...27M}, even when uncorrected
for cooling flows \citep{MCI}.  (2)~The temperature profiles of
the simulated clusters do not show the temperature decrement seen in the
central regions of real clusters \citep[e.g.][]{2001MNRAS.328L..37A}.

In order to lower the luminosity of the simulated
cluster enough to agree with observations, we can either adjust the
gravitational potential (by assuming a lower halo concentration)
or increase the entropy of the central gas. In order to get good agreement
with the average luminosities of clusters of this temperature, we would
need to adopt $c\sim3$, well below the concentrations inferred from 
dark matter
simulations \citep{2001ApJ...554..114E,2002ApJ...568...52W}. On the other hand,
the central entropy could be raised to $\ga 160$~keV~cm$^{-2}$
\citep{2002MNRAS.330..329B,2002ApJ...576..601V}, 
which is a factor of five greater than the initial entropy at 20~kpc.  Even if
enough energy were injected to achieve this (for example by running simulation
S2.0 for a much longer time), this would make the core temperature even
hotter, exacerbating the disagreement with observed temperature
profiles. Thus, we conclude that the initial cluster gas distribution
must have been different from the isothermal model we have assumed.  Since the
high cooling rate of the unperturbed cluster is a direct consequence of
its low central entropy (leading to high X-ray luminosity), our energy
injection mechanism has removed the symptoms of the overcooling
problem, but not addressed its cause.

Global preheating of the intragalactic medium prior to cluster
formation should lead to cluster gas distributions with integrated
luminosities and temperatures that are consistent with observations
\citep{1991ApJ...383..104K,1995MNRAS.275..720N,1999Natur.397..135P,1999MNRAS.307..463B,2002MNRAS.336..527M,2002MNRAS.336..409B}.
An alternative (but related) possibility is that heating events
similar to the ones we have simulated here occur in the cluster
progenitors and/or surrounding filamentary structure, at earlier
times.  Raising the entropy in these lower mass structures may result
in a higher entropy for the final cluster, as the lower density
results in an increase in the accretion shock strength
\citep{2003ApJ...593..272V,2003ApJ...594L..75V}.  In either scenario,
subsequent cooling would naturally lead to a temperature decrement in
qualitative agreement with observations.  Calculations by \citet{MCI}
suggest that although the cluster would then increase in luminosity,
it would decrease in emission weighted temperature so as to remain in
acceptable agreement with the observed L-T relation.

\subsection{Entropy transfer and generation}

In Figure~\ref{fig:energy}, we have shown that an energy injection
rate of $6.3\times 10^{44}\ergs$ is sufficient for the total energy of
the cluster to remain constant and for the density and temperature
distribution of the ICM to reach a quasi-equilibrium profile in which
the radiative heat losses are balanced by the energy deposited in the
ICM by the hot bubbles.  The slow evolution of the density profile
implies that the energy we inject is somehow shared with the cooling
ICM and does not remain trapped in the hot material. In particular,
the energy must be efficiently shared with the dense material near the
centre of the cluster, where the cooling time is short, leading to an
overall increase in entropy in these regions.

There are three mechanisms by which the energy can be shared:

\noindent(i) \textbf{$P\mbox{d}V$ work} -- as energy is injected into
the bubble, it expands doing work on the surrounding material. We
calculate the $P\mbox{d}V$ work by calculating the volume of the
bubble at a series of timesteps. $40$\% of the injected energy is used
in this way. The process is largely adiabatic, however, and has little
effect on the entropy distribution of the surrounding ICM.

\noindent(ii) \textbf{shocks} -- If the motion of the the bubble were
supersonic, shocks would dissipated energy in the surrounding ICM
leading to an overall increase in the entropy of the system. In
practice, however, the bubble does not reach supersonic speeds: as the
speed of the bubble increases the energy dissipation in the turbulent
wake grows. In our simulations the maximum Mach number in the flow is
0.8 thus we do not expect shocks to be a major source of entropy in
the surrounding ICM.

\noindent(iii) \textbf{up-lift and turbulence} -- The rising bubble
generates a complex flow pattern, with material from the central
regions of the cluster being up-lifted in the wake behind the
bubble. Vortices in this flow dissipate energy and mix the bubble
material, low entropy material from the center and the surounding ICM.
These irreversible processes generate an increase in the entropy of
the cluster.

Even though we assume that the viscosity of the ICM is small, it is
certainly not negligible, and viscous dissipation on molecular scales
can still be a major source of heating in turbulent regions of the
flow.  Clearly our simulations are far from resolving the relevant
molecular diffusion scales. However, in turbulent flows, the
non-linear couplings cause a cascade of energy from large to small
scale motions \citep{kolmogorov}.  The viscosity determines the
smallest scales in the flow, while the rate at which energy is
dissipated is determined by the largest scale eddies: these control
how quickly energy is fed into the turbulent cascade \citep{tennekes}.
In simulations, the development of the turbulent cascade is limited by
the numerical mixing of fluid elements within a grid cell, rather then
molecular viscosity.  However, the large scale properties of the flow
are not expected to be dependent on the smallest scales that are
resolved. We check this by comparing the evolution of a single bubble
in two simulations: namely the fiducial resolution and at almost an
order of magnitude greater linear resolution. Plotting the two
simulations to the same base resolution we see that the gross
structure of the flow is the same.  Thus, because of the weak
dependence of the large scale structure of the flow on the numerical
resolution, it is plausible that the energy dissipation and mixing
rates in the regions around the rising bubble will converge over the
range of resolution that we can test in Appendix~\ref{sec:conv}.

In addition to being a source of dissipative energy, the turbulence in
the flow provides a means of increasing the entropy of the surrounding
ICM through mixing.  The mixing may occur between the hot material
within the bubble and the surroundings, or between the surrounding ICM
and the low entropy material drawn out of the cluster centre by the
up-draft created by the rising bubble.  In both cases, the mixing is
irreversible and leads to an overall increase in the entropy of the
system.  In the first case, the mixing transfers energy from the bubble to
the surrounding ICM (in addition to the $P\mbox{d}V$ work discussed above).  
The second case does not directly transfer energy from the bubble, but by
driving an entropy increase in the lowest entropy material (the
material with the shortest cooling time) it prevents this material
cooling out of the flow.

\section{Conclusions}\label{sec:conc}

We have presented 3-dimensional gasdynamical simulations using the
{\sc Flash} adaptive--mesh refinement code of a cooling flow in an
isothermal, X-ray luminous cluster of galaxies, with periodic
injections of thermal energy. The initial cluster has mass of
$3\times10^{14}\Msol$ and a temperature of $3.1\keV$, consistent with
the observed mass--temperature relation \citep{2002ApJ...567..716R}.
The injected energy is manifested as hot bubbles of buoyant gas that
rise out of the cluster core, convectively mixing the cooling
material. We treat these injection events as sporadic outbursts with a
typical duty cycle of $100\Myr$. The parameters of these injection
events match the observed luminosity and frequency with which the most
powerful radio galaxies are seen in clusters.

In the absence of any energy injection, mass rapidly flows to the
cluster centre, at a rate that exceeds observational limits by at
least an order of magnitude.  Based on simulations with a variety of
parameters related to the energy injection rate, run for between 1.5
Gyr and 5 Gyr, we draw the following conclusions:

\begin{itemize}
\item[1.] For a time averaged energy injection rate of $6\times
  10^{44}\ergs$, the mass inflow rate is less than $30\Msolyr$,
  compatible with available observational limits.  With this amount of
  heat input, the total energy of the cluster remains approximately
  constant over 5 Gyr.
\item[2.] The evolution of the total cluster energy depends primarily
  on the total amount of energy injected and the spatial distribution
  of bubbles, but is only weakly sensitive to the duty cycle of
  heating events or to whether the bubbles are produced singly or in
  pairs.
\item[3.] The bubble activity generates concentric sound waves that are
  clearly evident in unsharp--masked projections of cluster emissivity.
\item[4.] When the injected energy just balances cooling, the entropy
  and temperature profile of the cluster remain approximately
  unchanged from their initial configurations.  
\end{itemize}

In summary, periodic energetic events of the kind we have simulated
can reduce the mass flow rate and accumulation of cold gas in massive
clusters to within observational limits.  However, this mechanism
operating on a fully formed cluster does not result in a final
luminosity consistent with observations.  It is likely that the
structure of the progenitors from which the cluster formed was
affected by heating events prior to the assembly of the final
cluster.

\section*{Acknowledgements}

We thank the anonymous referee for the stimulating questions he posed
in his report.  We are grateful to Romain Teyssier, Marcus Br\"uggen
and Alastair Edge for helpful discussions.  We thank Joop Schaye for
allowing us to use his cooling routines.  We thank Ian McCarthy for
providing his analytic cooling solutions prior to publication, and for
his helpful comments on this paper.  MLB and RGB acknowledge financial
support from PPARC fellowships, PPA/P/S/2001/00298 and
PPA/Y/S/2001/00407, respectively.  TT thanks PPARC for the award of an
Advanced Fellowship.  The software used in this work was in part
developed by the DOE-supported ASCI/Alliance Center for Astrophysical
Thermonuclear Flashes at the University of Chicago.

\newpage

\appendix

\section{Resolution Convergence Tests}\label{sec:conv}

\begin{figure}
\includegraphics[width=20pc]{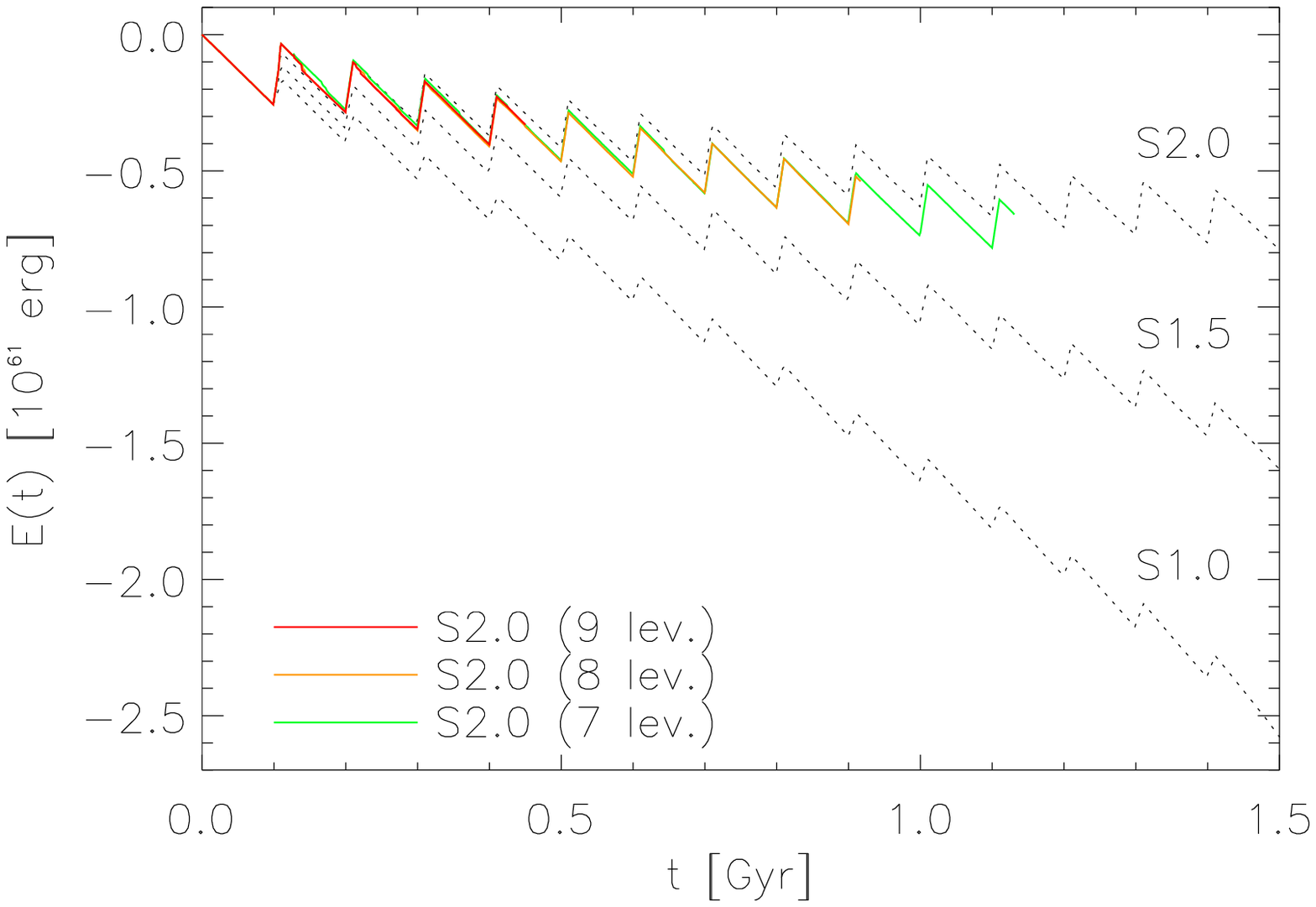}
\caption{As Fig.~\ref{fig:energy}, the energy evolution is shown for
  the simulations S1.0, S1.5 and S2.0 run with 6 levels of refinment
  (black, solid lines).  We compare these with simulation S2.0 run at
  different resolutions (7, 8 and 9 refinement levels).}
\label{fig:conv}
\end{figure}

We tested the convergence of the code by running simulation S2.0 at
increased resolution.  In Figure~\ref{fig:conv} we compare the total
energy evolution for simulation S1.0 (6 levels of refinement,
resolution of $7.6\kpc$) and S2.0 (6, 7 ($3.8\kpc$), 8 ($1.9\kpc$) and
9 ($0.9\kpc$) levels of refinement).  The codes were run for different
lengths of time because of the limitation of the computational time
and memory available. The differences between the four versions of the
S2.0 run are small in comparison to the relative evolution of the S1.5
and S2.0 simulations.  The figure also shows that the radiated energy
does not increase systematically with resolution. Comparing the total
radiated energy at $t=0.45\Gyr$, the four versions of the S2.0
simulation differ by less than 3\%.  At the last common output time,
the 6 and 7 level simulations differ by 4\%. We also compared the
density and entropy profiles of the simulations at this output time,
and found a similar level of agreement. We therefore chose to run the
code with a maximum refinement level of 6, this being a good
compromise between the speed of the code and the desired accuracy of
the relevant quantities.

Simulation S0.0 was also used to check the convergence of
energy conservation as the maximum number of levels of refinement
changes. With just 4 refinement levels, the total energy has an error
of $\sim 1\%$, while with 6 levels the error is less than $0.1\%$. We
decided to run the simulations with a maximum number of 6 levels. The
equivalent resolution of a fixed-grid Eulerian code is $256^3$.

\end{document}